\documentclass[pra, a4paper, twocolumn, 10pt]{revtex4}
\usepackage{bbm, amsmath, amssymb, amsthm, bm, textcomp, nicefrac}
\usepackage[utf8]{inputenc}
\usepackage[T1]{fontenc}
\usepackage[english]{babel}

\usepackage{graphicx}
\usepackage{geometry}
\usepackage{color}

\newcommand{\ket}[1]{| #1 \rangle}
\newcommand{\bra}[1]{\langle #1 |}
\newcommand*\colvec[3][]{ \begin{pmatrix}\ifx\relax#1\relax\else#1\\\fi#2\\#3\end{pmatrix} }

\theoremstyle{definition}

\theoremstyle{plain}

\begin{document}

\title{Effective noise channels for encoded quantum systems}

\author{Frederik Kesting, Florian Fr\"owis, Wolfgang D\"ur}
\affiliation{Institut f\"ur Theoretische Physik, Universit\"at  Innsbruck, Technikerstra\ss e 25, 6020 Innsbruck, Austria}
\date{\today}

\begin{abstract}  
  We investigate effective noise channels for encoded quantum systems with and without active error correction. Noise acting on physical qubits forming a logical qubit is thereby described as a logical noise channel acting on the logical qubits, which leads to a significant decrease of the effective system dimension. This provides us with a powerful tool to study entanglement features of encoded quantum systems. We demonstrate this framework by calculating lower bounds on the lifetime of distillable entanglement and the negativity for encoded multipartite qubit states with different encodings. At the same time, this approach leads to a simple understanding of the functioning of (concatenated) error correction codes.
\end{abstract}

\maketitle

\section{Introduction}\label{sec:introduction}
Quantum entanglement provides a valuable resource for many applications in quantum information science, ranging from quantum communication over quantum computation to quantum metrology. However, entanglement in multipartite states is generally fragile, where noise and decoherence diminish or destroy the desirable quantum features. Encoding quantum information redundantly is a possible approach to counter these effects. This can be done in an active way, where repetitive quantum error correction is used to actively detect and correct errors \cite{nielsen_quantum_2010}, or in a passive way where stabilization is obtained by using a certain subspace of a higher dimensional system which is less effected by noise.

In both cases, it is interesting to study the quantum features of encoded systems under the influence of noise (represented by quantum channels), in particular their entanglement properties and possible enhancements due to encoding. However, this is typically a complicated task, as the use of encodings significantly increases the dimension of the Hilbert space one needs to consider, thereby increasing the complexity of the problem at hand. Here, we discuss {\em effective noise channels} to simplify the description of the logical system in the presence of noise. The main idea is to derive an effective noise channel at the logical level, and use this to study the entanglement features of the logical multi-qubit state. That is, one considers encoded quantum information, where several physical qubits are used to encode one logical qubit. Errors acting on the individual qubits lead to the populations outside the logical subspace. However, active quantum error correction allows one to correct certain errors, while other errors may lead to an error at the logical level. This is done by performing syndrome measurements followed by appropriate correction operations. What is important in our context is that the system is effectively kept within the two-dimensional logical subspace. This implies that one can derive an effective error channel at the logical level, allowing one to describe the total system as one logical qubit. The merit of a certain encoding is then conveyed to the corresponding effective noise channel, and entanglement features of encoded multipartite states can be obtained by studying the multi-qubit system under the influence of different (effective) noise channels. In particular, this allows one to use results obtained in the study of unencoded systems.

Similarly, also in the case of passive protection, one can use such an approach to study different entanglement features of encoded multipartite entangled states under the influence of noise. This can be done by considering for each logical system either (i) only the logical subspace and deriving an effective (renormalized) noise channel within this subspace, or (ii) by splitting the total Hilbert space of the logical system in orthogonal two-dimensional subspaces that can be treated independently. In both cases, effective noise channels can be derived and used to establish entanglement features of encoded systems. In the case of (i), one can, e.g., establish lower bounds on the lifetime of distillable entanglement, while the approach (ii) can be used to derive lower bounds on different entanglement measures, e.g., the negativity of entanglement \cite{peres_separability_1996,vidal_computable_2002}. The second approach can be further simplified by considering a {\em mean} noise channel, still leading to lower bounds for entanglement measures due to the fact that the averaging corresponds to a local operation (in terms of the logical system) that can only diminish entanglement.

Here, we derive a framework for effective noise channels and apply it to different encodings. In the case of an optimal five-qubit error correction code \cite{laflamme_perfect_1996,bennett_mixed-state_1996,schlingemann_quantum_2001,grassl_graphs_2002,schlingemann__2002}, we find that depolarizing noise at the physical level leads to effective depolarizing noise at the logical level, where the noise parameter is decreased as long as the initial noise is sufficiently weak. For a repetition code 
capable of correcting bit-flip errors, we find that depolarizing noise at the physical level leads to effective Pauli noise with a preferred direction at the logical level. That is, while logical phase-flip errors are slightly enhanced, logical bit-flip and joint phase-flip and bit-flip errors are exponentially suppressed. This allows one to understand the encountered stability of so-called concatenated Greenberger-Horne-Zeilinger (GHZ) states \cite{frowis_stable_2011} --logical GHZ states where GHZ states are used for passive encoding-- in a simple way: At the logical level, the resulting effective noise has a preferred direction, and GHZ states in a certain basis show a significantly enhanced robustness under such noise \cite{chaves_robust_2012} as compared to (undirected) depolarizing noise or phase noise. In a sense, the encoding only transforms the noise to a form the system can better cope with. 

With the help of such effective noise channels, also concatenated error correction codes can be easily analyzed. Once an effective noise channel for a given error model has been derived, it can be used to obtain the effective noise channel for a concatenated error correction code by a sequential application of the corresponding maps. Also the effect of different codes at different concatenation levels can be easily taken into account.
 
This paper is organized as follows. In Sec.~\ref{sec:effective-noise-description}, we describe how to obtain effective noise channels for passive error protection and active error correction. In Sec.~\ref{sec:effect-noise-repet}, we explicitly derive and analyze effective noise channels for repetition and cluster-ring encodings of different sizes. In Sec.~\ref{sec:entanglement-measures}, we apply these effective noise channels and obtain lower bounds on the lifetime of distillable entanglement and the negativity of entanglement for encoded GHZ states, and compare the effect of different encodings. In Sec.~\ref{sec:conc-encod}, we discuss concatenated encodings and potential applications in the context of quantum error correction, while we summarize and conclude in Sec.~\ref{sec:outlook}.

\section{Effective Noise Description}\label{sec:effective-noise-description}

In this section, we review the basic concept of logical encoding and quantum error correction and present --as one of our main results-- a formalism for the effective description of noise on a logical level.

The basic idea of quantum error correction is to encode the information in a higher-dimensional space. Typically, one identifies \textit{physical} entities as qubits, which are the basic quantum information units defined on $\mathcal{H}_1 =\mathbbm{C}^{2}$. With these physical qubits, one builds a \textit{logical} qubit. The Hilbert space of $m$ physical qubits is $\mathcal{H}_m = \mathbbm{C}^{2 \otimes m}$. The logical qubit is defined by specifying a two-dimensional subspace in $\mathcal{H}_m$, which is here denoted by $P_0$. Therefore, one chooses a map
\begin{equation}
 \label{eq:1}
 \begin{split}
  \left| 0 \right\rangle \in \mathcal{H}_1 & \mapsto \left| 0 _{\mathrm{L}}\right\rangle \in \mathcal{H}_m,\\
  \left| 1 \right\rangle \in \mathcal{H}_1 & \mapsto \left| 1 _{\mathrm{L}}\right\rangle \in \mathcal{H}_m,
 \end{split}
\end{equation}
where $\left\{ | 0 \rangle, | 1 \rangle  \right\}$ are an orthonormal (ON) basis of $\mathcal{H}_1$ and $\left\{ | 0_{\mathrm{L}} \rangle, | 1_{\mathrm{L}} \rangle  \right\}$ are an ON basis of $P_0$. The remaining Hilbert space is divided into orthogonal, two-dimensional subspaces $P_i, i \in \left\{ 1,\dots,2^{m-1}-1 \right\}$. This is ideally done such that typical errors on the physical qubits (e.g., Pauli errors on one or several qubits) map the space $P_0$ to distinct spaces $P_i$. As an instance, consider an optimal five-qubit error correction code. The Hilbert space of five qubits is 32-dimensional, which implies a partitioning into 16 subspaces. One subspace is the logical subspace $P_0$, the remaining 15 subspaces constitute from three possible errors per qubit. The errors are phase-flip (Z), bit-flip (X) and joint phase-flip and bit-flip (Y). The map (\ref{eq:1}) has to be chosen such that X, Y and Z errors map $P_0$ to distinct orthogonal subspaces $P_{i>0}$. The code is called optimal, because five qubits are the minimal number of qubits that is capable to correct one arbitrary single-qubit error.

The standard procedure to correct errors on the logical qubit is to first measure the logical qubit with $\{P_{i}\}_{i = 0}^{2^{m-1}-1}$ as the eigenspaces of the measurement. This reads out the ``syndrome'' and projects the quantum state onto one of the spaces $P_i$. Next, one corrects the error by applying a unitary operation that maps $P_i$ back to $P_0$.

In most cases, one has to deal with more errors than available subspaces. Then, this procedure can in general not correct any error, but only a subset. The remaining errors lead then to a \textit{logical} error. However, a successful code (i.e., a good partitioning of $\mathcal{H}_m$) reduces the effective error on the logical level compared to the possible errors on one physical qubit.

The theory of error correction is important in many quantum information applications. For instance, in the context of quantum computing, noise can destroy coherences within the total quantum state. This can quickly diminish the success rate of an algorithm to a level where no improvements compared to classical algorithms can be expected. However, the idea of an (redundant) encoding of physical qubits in logical qubits can be used to protect quantum properties of many-body quantum states in general. In particular, we are here interested in the entanglement properties of multipartite qubit states when the physical qubits are encoded into logical qubits. Consider a quantum state $| \psi \rangle \in \mathcal{H}_1^{\otimes N}, N \in \mathbbm{N}$. Under the map (\ref{eq:1}), $| \psi \rangle$ is mapped to $ | \psi_{\mathrm{L}} \rangle \in \mathcal{H}_m^{\otimes N} = \mathcal{H}_1^{\otimes Nm}$. We now ask how entanglement alters under the effect of local noise. If a single qubit channel is denoted by $\mathcal{E}: \mathcal{B}(\mathcal{H}_1)\rightarrow \mathcal{B}(\mathcal{H}_1)$ [$\mathcal{B}(\mathcal{H})$ is the set of density operators on $\mathcal{H}$], the total action results in $\rho_{\mathrm{L}} = \mathcal{E}^{\otimes Nm} (|\psi_{\mathrm{L}}  \rangle\!\langle \psi_{\mathrm{L}}|)$. It is interesting to study which encodings [i.e., which maps as in Eq.~(\ref{eq:1})] increase the resistance of $| \psi_{\mathrm{L}} \rangle $ in the presence of noise; with or without the active correction procedure consisting of measurement and correcting unitary. As simple examples show (see Sec.~\ref{sec:effect-noise-repet}), even the choice of the basis within $P_0$ can change the entanglement properties of the total state.

It is crucial to understand that entanglement properties of an encoded quantum state can be protected through the encoding (\ref{eq:1}) alone. The subsequent active correction procedure --composed of a measurement on the logical qubit and potential correction-- is a local operation on the logical level. It can therefore not increase entanglement between the blocks. In order to analyze the merit of a particular encoding for increased stability, it is therefore not necessary to actually simulate the error correction. However, the complexity of encoding a qubit within $\mathcal{H}_m$ renders the actual computation of the stability exponentially difficult in $m$. Here, the correction procedure can help to reduce this complexity, because the syndrome  measurement projects the high-dimensional reduced state of one logical block onto a two-dimensional subspace. As we will see in the following, one can replace the noise map $\mathcal{E}^{\otimes m}$ by an effective map $\tilde{\mathcal{E}}: \mathcal{B}(P_0) \rightarrow \mathcal{B}(P_0)$. (Notice that $P_0$ is isomorphic to $\mathcal{H}_1$.) This reduces the computational complexity drastically, since one can treat $| \psi_{\mathrm{L}} \rangle $ as an effective $N$ qubit state without encoding. The merit of the encoding is imprinted in the modified map $\tilde{\mathcal{E}}$. Then, one may use already known results for $| \psi \rangle $ for studies of $| \psi_{\mathrm{L}} \rangle $. Notice, however, that the entanglement properties of $\tilde{\mathcal{E}}^{\otimes N}(|\psi  \rangle\!\langle \psi|)$  are only lower bounds on those of $\mathcal{E}^{\otimes Nm}(|\psi_{\mathrm{L}}  \rangle\!\langle \psi_{\mathrm{L}}|)$, since it emerged from local operations (projections onto $P_i, i\geq 0$).

In the remainder of this section, we will present two possibilities to derive two different kinds of $\tilde{\mathcal{E}}$ with different fields of application. The proposed methods are generally applicable. However, we demonstrate them for two restrictions.

First, we consider the so-called Pauli noise channel. For any $\rho \in \mathcal{B}(\mathcal{H}_1)$, this single-qubit channel can be written as
\begin{equation}\label{eq:initial-noise}
 \mathcal{E}(\rho) = \sum_{j=0}^3\lambda_j\sigma_j\rho\sigma_j,
\end{equation}
with $\sigma_j \in \left\{ \sigma_0 \equiv \mathit{id}_{\mathcal{H}_1}, \sigma_1 \equiv \sigma_x, \sigma_2 \equiv \sigma_y, \sigma_3\equiv \sigma_z \right\}$ denoting the identity operator and the Pauli operators and the noise parameters $\lambda_j \in [0,1]$ such that $\sum_{j = 0}^3 \lambda_j = 1$. This class of noise channels includes the so-called depolarization channel or white noise for the parameter choice
\begin{equation}
 \lambda_0 = \frac{1 + 3p}{4}, \quad
 \lambda_{j>0} = \frac{1-p}{4} \label{eq:13}
\end{equation}
with $p \in [0,1]$. White noise is undirected in the sense that it is invariant under single-qubit unitaries. Another important instance is given through $\lambda_0 = (1+p)/2$ and $\lambda_3 = (1-p)/2$ implying $\lambda_1 = \lambda_2 = 0$. This choice represents phase noise. Using the standard convention that $\left\{ \left| 0 \right\rangle , \left| 1 \right\rangle  \right\}$ is the eigenbasis of $\sigma_z$, this set of states is invariant under the phase noise channel.

Second, we focus on stabilizer states \cite{gottesman_class_1996} as the logical states $| 0_{\mathrm{L}} \rangle $ and $| 1_{\mathrm{L}} \rangle $ from Eq.~(\ref{eq:1}). Stabilizer states constitute an important class of quantum states for quantum error correction and in fact many of the most well known error codes are stabilizer codes \cite{nielsen_quantum_2010,laflamme_perfect_1996,bennett_mixed-state_1996,shor_scheme_1995}. A stabilizer state $| \mathrm{S} \rangle \in \mathcal{H}_m$ is uniquely defined as the eigenstate of $2^m$ operators of the so-called Pauli group \footnote{The Pauli group is generated by the operators $\sigma_i^{(j)} \equiv \sigma_0^{\otimes j-1} \otimes \sigma_i \otimes \sigma_0^{\otimes n-j-1}$.} with eigenvalue one. An important subclass of stabilizer states are 
graph states \cite{briegel_persistent_2001,hein_entanglement_2005-1}. For the definition of a graph state, consider a simple graph where the vertices represent the qubits and the edges defines the neighborhood $N_a$ of any vertex $a$. A graph state $| \mathrm{G} \rangle $ is then defined as the unique state that is an eigenstate of
\begin{equation}
 K_a = \sigma_x^{(a)}\prod_{j \in N_a}\sigma_z^{(j)}\label{eq:14}
\end{equation}
with eigenvalue $+1$, $a \in \left\{ 1,\dots,m \right\}$. Here, we identify $| 0_{\mathrm{L}} \rangle =| \textrm{G} \rangle $ and $| 1_{\mathrm{L}} \rangle = \sigma_z^{\otimes m} | \textrm{G} \rangle $ (i.e., $| 1_{\mathrm{L}} \rangle$ is the unique eigenstate of all $K_a$ with eigenvalue $-1$). Graph states are equivalent to stabilizer states up to local Clifford operators \cite{hein_entanglement_2005-1}.

With these two restrictions, we assume in the following that the $P_{i>0}$ are constructed by applying certain elements of the Pauli group on $P_0$. These operators span the space of the most likely errors.

\subsection{Projection onto $P_i$}
\label{sec:projection-p_0}

For qualitative questions on entanglement (i.e., whether the noisy, encoded state is entangled or not), it suffices to apply \textit{stochastic} operations locally on the level of logical blocks. If we find entanglement in the state after such an operation, the state was entangled already before. To develop the basic idea, consider one logical block. For an arbitrary state $\rho \in \mathcal{B}(P_0)$, one first applies the physical noise $\mathcal{E}^{\otimes m}$ which in general alters the support to the entire $\mathcal{H}_m$. Then, one projects onto one of the subspaces $P_i$. After renormalization, the new state is rotated back to the space $P_0$. Since the outcome of this procedure is a quantum state in $\mathcal{B}(P_0)$, it can be expressed as 
\begin{equation}\label{eq:eff-noise}
\tilde{\mathcal{E}}^{(i)}(\rho) = \sum_{j,k=0}^3 \tilde{\lambda}^{(i)}_{j,k}\sigma_j^L\rho\sigma_k^L.
\end{equation}
The operators $\sigma_j^L: P_0 \rightarrow P_0$ are \textit{logical} Pauli operators that are defined for $\left\{ \left| 0_{\mathrm{L}} \right\rangle , \left| 1_{\mathrm{L}} \right\rangle  \right\}$ analogously as the standard Pauli operators for $\left\{ \left| 0 \right\rangle , \left| 1 \right\rangle  \right\}$. The parameters $\tilde{\lambda}^{(i)}_{j,k}$ are noise parameters of the corresponding effective noise channels and depend on the choice of the map (\ref{eq:1}), on $\lambda_i$ from the physical Pauli channel and on the space $P_i$ one projects on. By construction, the map $\tilde{\mathcal{E}}_i$ is a completely positive (cp), trace preserving map \cite{nielsen_quantum_2010}.

We derive the effective noise description (\ref{eq:eff-noise}) in a simple way by making use of the Choi-Jamio{\l}kowski Isomorphism \cite{de_pillis_linear_1967,jamiolkowski_linear_1972} between cp maps and higher-dimensional quantum states. We therefore define the maximally entangled state
\begin{equation}
 \ket{\Phi^+} = \frac{1}{\sqrt{2}}\left(\ket{0_{\mathrm{L}}}\otimes\ket{0}+\ket{1_{\mathrm{L}}}\otimes\ket{1} \right) \in P_0 \otimes \mathcal{H}_1.
\end{equation}
The state resulting from the action of $\mathcal{E}^{\otimes m} \otimes \mathit{id}_{\mathcal{B}(\mathcal{H}_1)}$ on $| \Phi^{+} \rangle\!\langle \Phi^{+} |$  with subsequent projection onto any $P_i \otimes \sigma_0$ and unitary rotation to $P_0$ is isomorphic to $\tilde{\mathcal{E}}^{(i)}$ from Eq.~(\ref{eq:eff-noise}).

For any stabilizer state, the following pattern emerges from this procedure \footnote{The coefficients $x_k, x\in \left\{ a,b,c,d \right\}$ and $k \in \left\{ 0,1 \right\}$ depend on $i$; but for simplicity, this dependency is suppressed in the notation if possible.}:
\begin{equation*}
  \begin{split}
    \ket{0_{\mathrm{L}}}\bra{0_{\mathrm{L}}}\otimes\ket{0}\bra{0} &\mapsto \left(
      a_0\ket{0_{\mathrm{L}}}\bra{0_{\mathrm{L}}}
      +b_0\ket{1_{\mathrm{L}}}\bra{1_{\mathrm{L}}}\right)\otimes\ket{0}\bra{0} \\
    \ket{1_{\mathrm{L}}}\bra{1_{\mathrm{L}}}\otimes\ket{1}\bra{1} &\mapsto  \left(
      a_1\ket{1_{\mathrm{L}}}\bra{1_{\mathrm{L}}}
      +b_1\ket{0_{\mathrm{L}}}\bra{0_{\mathrm{L}}}\right)\otimes\ket{1}\bra{1} \\
    \ket{0_{\mathrm{L}}}\bra{1_{\mathrm{L}}}\otimes\ket{0}\bra{1} &\mapsto  \left(
      c_0\ket{0_{\mathrm{L}}}\bra{1_{\mathrm{L}}}
      +d_0\ket{1_{\mathrm{L}}}\bra{0_{\mathrm{L}}}\right)\otimes\ket{0}\bra{1} \\
    \ket{1_{\mathrm{L}}}\bra{0_{\mathrm{L}}}\otimes\ket{1}\bra{0} &\mapsto  \left(
      c_1\ket{1_{\mathrm{L}}}\bra{0_{\mathrm{L}}}
      +d_1\ket{0_{\mathrm{L}}}\bra{1_{\mathrm{L}}}\right)\otimes\ket{1}\bra{0}
  \end{split}
\end{equation*}
For symmetry reasons one easily sees that $a_0=a_1  \equiv a$, $b_0=b_1  \equiv b$, $c_0=c_1 \equiv c$ and $d_0=d_1 \equiv d$. In the basis spanned by the set $\left\{ \left| i_{\mathrm{L}} \right\rangle \otimes \left| j \right\rangle  \right\}_{i,j = 0}^1$, this results in the matrix
\begin{equation}\label{eq:matrix-eff-noise}
 M = \frac{1}{2}\left( \begin{array}{cccc}
 a & 0 & 0 & c \\
 0 & b & d & 0 \\
 0 & d & b & 0 \\
 c & 0 & 0 & a \end{array} \right).
\end{equation}
We observe that $M$ is diagonal in the Bell basis. The resulting effective noise description for Pauli noise on encodings is thus again Pauli noise. The effective noise parameters $\tilde{\lambda}^{(i)}_{j,k}$ can be deduced by transforming Eq. (\ref{eq:matrix-eff-noise}) into the Bell basis and renormalization:
\begin{equation}
 \begin{aligned}
  \tilde{\lambda}^{(i)}_{0} \equiv \tilde{\lambda}^{(i)}_{0,0} &= \frac{a+c}{2a+2b}, & \tilde{\lambda}^{(i)}_{1} \equiv\tilde{\lambda}^{(i)}_{1,1} &= \frac{b+d}{2a+2b}, \\
  \tilde{\lambda}^{(i)}_{2} \equiv  \tilde{\lambda}^{(i)}_{2,2} &= \frac{b-d}{2a+2b}, & \tilde{\lambda}^{(i)}_{3} \equiv \tilde{\lambda}^{(i)}_{3,3} &=  \frac{a-c}{2a+2b}, 
 \end{aligned}\label{eq:2}
\end{equation}
and $\tilde{\lambda}^{(i)}_{j,k} = 0$ for $j \neq k$.
With these \textit{effective} channels at hand, we can identify the subspace $P_i$ where entanglement persists longest (typically it is $P_0$) and calculate the entanglement lifetime of $\tilde{\mathcal{E}}^{(0) \otimes N} (\left| \psi_{\mathrm{L}} \right\rangle\!\left\langle \psi_{\mathrm{L}}\right| )$, which is computationally similar to the original problem $\mathcal{E}^{\otimes N} (\left| \psi \right\rangle\!\left\langle \psi\right| )$, but gives a lower bound on the entanglement lifetime of $\mathcal{E}^{\otimes Nm} (\left| \psi_{\mathrm{L}} \right\rangle\!\left\langle \psi_{\mathrm{L}}\right| )$.

In case one is interested in estimating entanglement of an encoded, noisy state quantitatively, one has to take into account \textit{all} possible outcomes of the syndrome measurement. Considering $N$ logical blocks and applying the techniques of this section, one has $2^{N(m-1)}$ different configurations $\tilde{\mathcal{E}}^{(i_1)}\otimes \dots \otimes \tilde{\mathcal{E}}^{(i_N)}$ ($i_k \in \{0,\dots, 2^{m-1}-1\}$). This exponentially large number (in $m$ and $N$) renders the treatment for such questions inefficient. Therefore, we consider a simplification of the problem in the next section.

\subsection{Mean Noise Channels}
\label{sec:aver-noise-chann}

We here introduce a \textit{mean noise channel} $\tilde{\mathcal{E}}_{\mathrm{mean}}(\rho) $  that can be used to estimate the decoherence of the logical qubit state. The idea is to perform the measurement with projection operators onto the set $\left\{ P_i \right\}_{i = 0}^{2^m-1}$, make the correction operation back to the $P_0$ and treat all outcomes on equal footing (i.e., we average over all possible outcomes). For any $\rho\in \mathcal{B}(P_0)$, the mean noise channel is defined as
\begin{equation}\label{eq:4}
 \tilde{\mathcal{E}}_{\mathrm{mean}}(\rho) = \sum_{j,k=0}^3\mu_{j,k}\sigma_j^L\rho\sigma_k^L.
\end{equation}
The parameters  $\mu_{j,k}$ are mean noise parameters defined as the sum of the effective noise parameters $\tilde{\lambda}_{j,k}^{(i)}$ weighted by the probability $p_i = a^{(i)} + b^{(i)}$ to find $\mathcal{E}^{\otimes m} (\rho)$ in the subspace $P_i$, that is,
\begin{equation}\label{eq:5}
 \mu_{j,k} = \sum_{i=0}^{2^{m-1}-1}p_i\tilde{\lambda}_{j,k}^{(i)}.
\end{equation}
Again, for stabilizer codes and Pauli noise as the physical noise source, the ``off-diagonal'' elements of the mean noise channel vanish, that is, $\mu_{j,k} = 0$ for $j \neq k$. Hence, the mean noise channel has the form of a Pauli channel.

Quantifying the entanglement of $\tilde{\mathcal{E}}_{\mathrm{mean}}^{\otimes N}(\left| \psi_{\mathrm{L}} \right\rangle\!\left\langle \psi_{\mathrm{L}}\right| )$ with any entanglement measure gives us lower bounds on the entanglement of $\mathcal{E}^{\otimes Nm}(\left| \psi_{\mathrm{L}} \right\rangle\!\left\langle \psi_{\mathrm{L}}\right| )$, because the projection and summation in Eq.~(\ref{eq:5}) is a local operation and entanglement cannot increase under those.

\section{Examples: Repetition and Cluster-Ring Code}
\label{sec:effect-noise-repet}

In this section, examples of effective noise channels for two different choices of encoding are derived. First, we investigate the code space spanned by
\begin{equation}
 \label{eq:9}
 \begin{split}
  \left| 0_{\mathrm{L}} \right\rangle &\doteq  
      \left| 0 \right\rangle ^{\otimes m} ,\\
    \left| 1_{\mathrm{L}} \right\rangle &\doteq   \left| 1 \right\rangle
      ^{\otimes m},
 \end{split}
\end{equation}which is a simple repetition code to correct bit-flip errors. Second, we consider a five-qubit code that can correct one arbitrary error on a physical qubit \cite{laflamme_perfect_1996,bennett_mixed-state_1996,schlingemann_quantum_2001,grassl_graphs_2002,schlingemann__2002}. We represent the latter in terms of a graph state, in particular as a so-called one-dimensional cluster state \cite{briegel_persistent_2001} with periodic boundary conditions. We refer to this code as the cluster-ring encoding. Note that both codes are stabilizer codes.

We start with the repetition code. This code can correct up to $\lfloor m/2 \rfloor$ bit-flip errors. The subspaces $P_i$ can be constructed by applying all possible combinations of up to $\lfloor m/2 \rfloor$ bit-flip operations on the logical subspace $P_0$. For a fixed number $i$ of errors, there are $\binom{m}{i}$ different possibilities to distribute $i$ errors among $m$ qubits. By applying the protocol presented in Sec.~\ref{sec:projection-p_0}, we find an analytical expression for $\tilde{\mathcal{E}}_{i}$ for all $i$ and $m$ \footnote{Note the slight abuse of notation, since $i\in \left\{ 0,\dots,\lfloor m/2 \rfloor \right\}$ indicates the number of errors.}. The matrix elements of Eq.~(\ref{eq:matrix-eff-noise}) read
\begin{equation}
 \label{eq:8}
 \begin{split}
  a^{(i)} &= \frac{1}{2}\left( \lambda_0+\lambda_3\right)^{m - i} \left(\lambda_1 + \lambda_2 \right)^i\\
  b^{(i)} &= \frac{1}{2}\left( \lambda_0+\lambda_3\right)^{i} \left(\lambda_1 +  \lambda_2 \right)^{m-i}\\
  c^{(i)} &= \frac{1}{2}\left( \lambda_0-\lambda_3\right)^{m - i} \left(\lambda_1 - \lambda_2 \right)^i\\
  d^{(i)} &= \frac{1}{2}\left( \lambda_0-\lambda_3\right)^{i} \left(\lambda_1 -  \lambda_2 \right)^{m-i}\\
 \end{split}
\end{equation}
Using Eq.~(\ref{eq:2}), one easily gets the expressions for $\lambda_k^{(i)}$. For weak depolarizing noise (\ref{eq:13}) on the physical level (i.e., $p = 1-\epsilon, \epsilon \ll 1$), we approximate these expressions to gain some insight. For instance, the expressions for $k = 0$ (i.e., no error) are
\begin{equation}
 \begin{aligned}
  \tilde{\lambda}_0^{(0)} & \approx 1 - \frac{m \epsilon}{4}, &
  \tilde{\lambda}_1^{(0)} &\approx \frac{\epsilon^m}{2^{m+1}},\\
  \tilde{\lambda}_2^{(0)} &  \approx \frac{\epsilon^m}{2^{m+1}}, &
  \tilde{\lambda}_3^{(0)} & \approx   \frac{m \epsilon}{4}.
 \end{aligned}
\end{equation}
We see that the effective X and Y noise is exponentially damped with $m$, while the effective Z noise is linearly increased by $m$. This naturally reflects the fact that the repetition code can correct X errors but is increasingly sensitive to Z errors. Physical Y errors are only corrected partially. The ``flip'' part is corrected, the ``phase'' part cannot be corrected and a Y error therefore counts as a Z error on the logical level.

\begin{figure}[htbp]
\centerline{\includegraphics[width=\columnwidth]{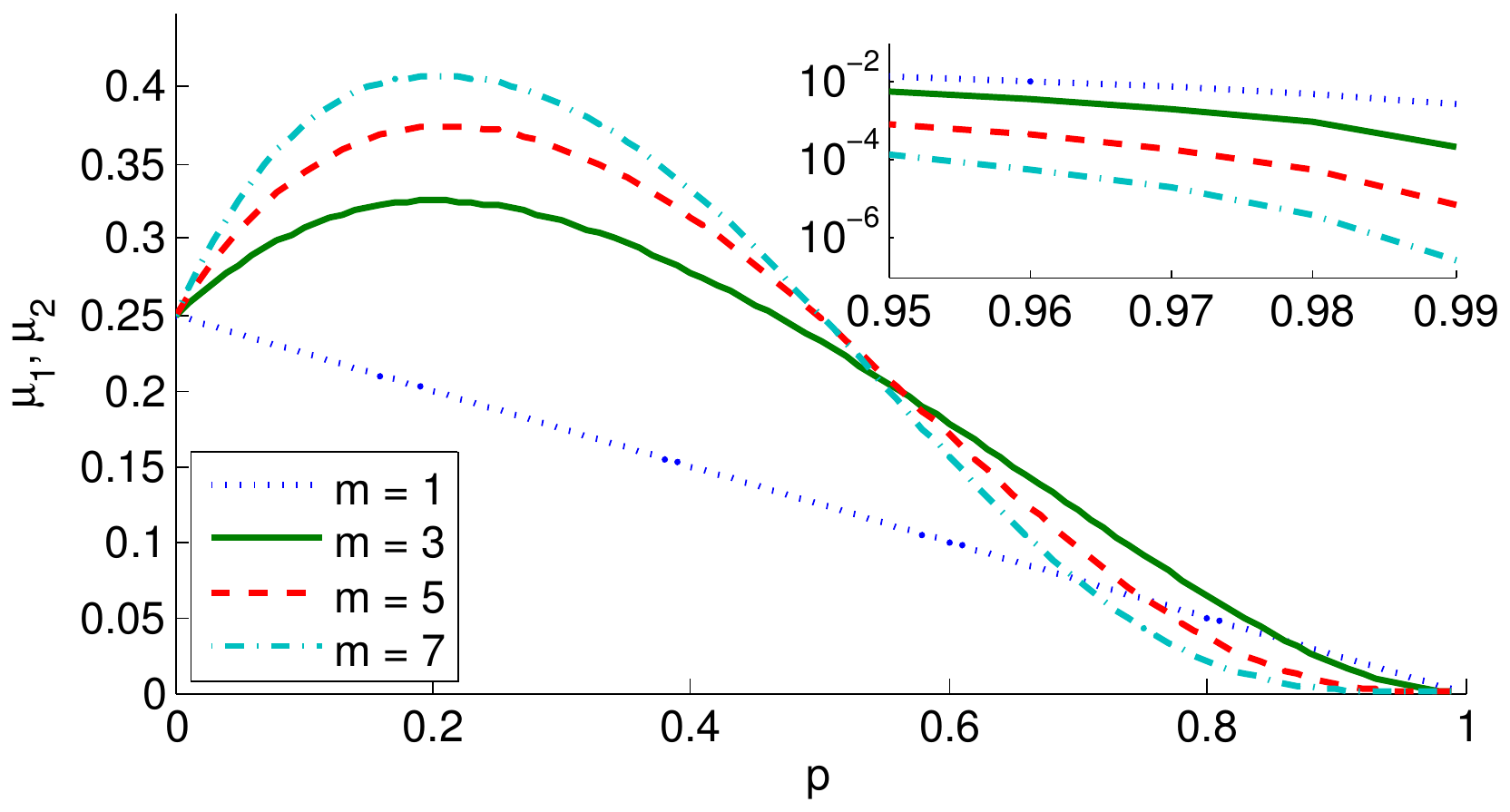}}
\caption[]{\label{fig:GHZ_XYNoise} Mean X and Y noise parameters $\mu_1 = \mu_2$ for repetition code and physical white noise (\ref{eq:13}) for different noise parameters $p$. The important parameter regime is for $p$ close to one, where a decrease of the effective noise rates can be achieved compared to the physical noise (which is identical to $m = 1$). In particular, a larger group size $m$ results in an exponential stabilization with respect to X errors. This can be seen from the inlet, which is a logarithmic plot for weak noise (i.e., for $p$ close to one).}
\end{figure}

\begin{figure}[htbp]
\centerline{\includegraphics[width=\columnwidth]{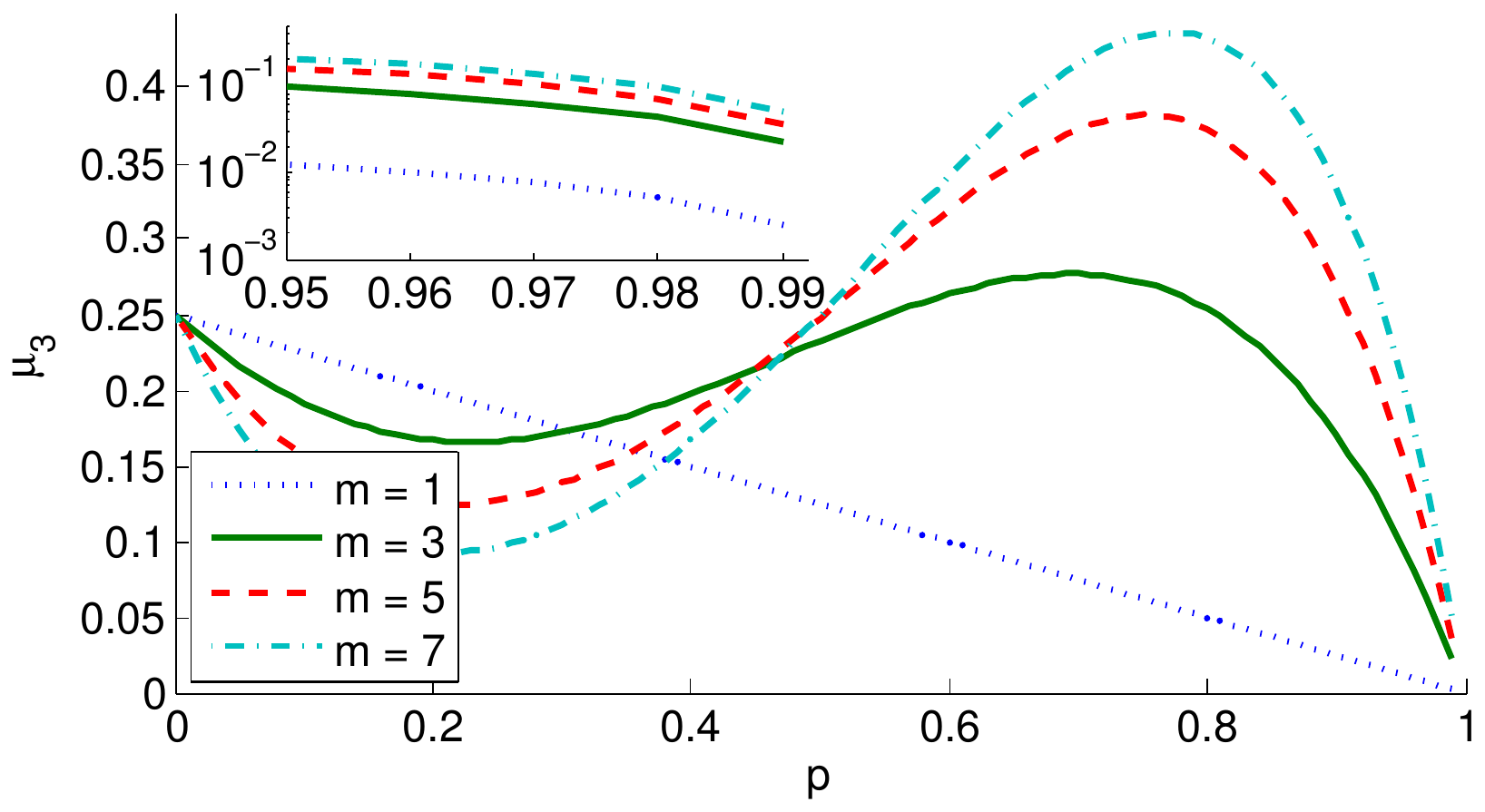}}
\caption[]{\label{fig:GHZ_ZNoise} Mean Z noise parameter $\mu_3$ for repetition code and physical white noise for different noise parameters $p$. Increasing the number $m$ of physical qubits per logical block increases the sensitivity to phase noise, since the code cannot correct phase-flip errors. The inlet shows part of the data in a logarithmic plot.}
\end{figure}

For the mean noise channel (\ref{eq:4}), one has to sum up the contributions for different $i$. One easily finds
\begin{equation}
\label{eq:11}
\begin{split}
  \mu_0 &= \frac{1}{2} \sum_{i = 0}^{\lfloor m/2 \rfloor}\binom{m}{i} \left( a^{(i)} + c^{(i)} \right),\\
  \mu_1 &= \frac{1}{2} \sum_{i = 0}^{\lfloor m/2 \rfloor}\binom{m}{i} \left( b^{(i)} + d^{(i)} \right),\\
  \mu_2 &= \frac{1}{2} \sum_{i = 0}^{\lfloor m/2 \rfloor}\binom{m}{i} \left( b^{(i)} - d^{(i)} \right),\\
  \mu_3 &= \frac{1}{2} \sum_{i = 0}^{\lfloor m/2 \rfloor}\binom{m}{i} \left( a^{(i)} - c^{(i)} \right).
\end{split}
\end{equation}
For small noise, the effective noise parameters $\mu_1$ and $\mu_2$ (i.e., flip and phase-flip channels) are exponentially suppressed with increasing $m$, while the phase error represented by $\mu_3$ is increased (see Figs.~\ref{fig:GHZ_XYNoise} and \ref{fig:GHZ_ZNoise}).

As an alternative, we consider the cluster-ring encoding based on the choice 
\begin{equation}\label{eq:cr-enc}
 \begin{split}
  \left| 0_{\mathrm{L}} \right\rangle \doteq \ket{\mathrm{Cl}_m^+} &= \prod_{i=1}^mC^{(i,i+1)}\ket{+}^{\otimes m}\\
  \left| 1_{\mathrm{L}} \right\rangle \doteq \ket{\mathrm{Cl}_m^-} &= \prod_{i=1}^mC^{(i,i+1)}\ket{-}^{\otimes m},
 \end{split}
\end{equation} where $C=\left| 0 \right\rangle\!\left\langle 0\right| \otimes \sigma_0 +\left| 1 \right\rangle\!\left\langle 1\right| \otimes\sigma_3$ denotes the phase gate. It is applied on all neighboring qubits $(i,i+1)$ with $m+1 \equiv 1$. For $m=5$, this code is able to correct exactly one error on any physical qubit. Although some of the following results are valid for any $m$, we restrict ourselves in the remainder of this section to $m=5$.

The subspaces $P_i$ are constructed by applying all possible combinations of $\sigma_z$ operators on the logical space $P_0$ with the condition that the total number of $\sigma_z$ is at most $\lfloor m/2 \rfloor$. For $m=5$, this is compatible with the construction of the $P_i$ for the optimal five-qubit code as described in Sec.~\ref{sec:effective-noise-description}: An X error on qubit $a$ corresponds to Z errors on the qubits $a-1$ and $a+1$. A Y error on $a$ translates to Z errors on the qubits $a-1, a$ and $a+1$ (consider the action of $\sigma_x^{(a)}$ on $K_{a}$ and note that the code words (\ref{eq:cr-enc}) are eigenstates of $K_a$ \cite{hein_entanglement_2005-1}). This indeed gives rise to 15 different combinations of products of local $\sigma_z$ operations. 

The calculation of the effective noise coefficients is not as straightforward as in the case of the repetition code. We focus therefore on white noise as the physical noise source and $m=5$. One finds that the physical white noise is transformed to white noise on the logical level. The effective noise parameter for the subspace $P_0$ is $\tilde{p}^{(0)} \mathrel{\mathop:}= (4 \tilde{\lambda}_0^{(0)} - 1)/4$ and reads
\begin{equation}
 \label{eq:12}
 \tilde{p}^{(0)} = \frac{1 -10 x^3 + 15 x^4 - 6 x^5}{1 + 30 x^3 + 15 x^4 + 18 x^5} \approx 1 - \frac{5}{8}(1-p)^3,
\end{equation}
with $x = (1-p)/(1+3p)$. The approximation is valid for $p$ close the one. This means that in the weak noise regime and in this subspace, the effective noise parameter is suppressed with the third power.
We numerically find that the effective noise parameters for a cluster-ring encoding in the logical subspace is again white noise for all $m$. Compared to the physical noise, the effective error rates are reduced in the weak noise regime.

\begin{figure}[htb]
 \center
 \includegraphics[width= \columnwidth]{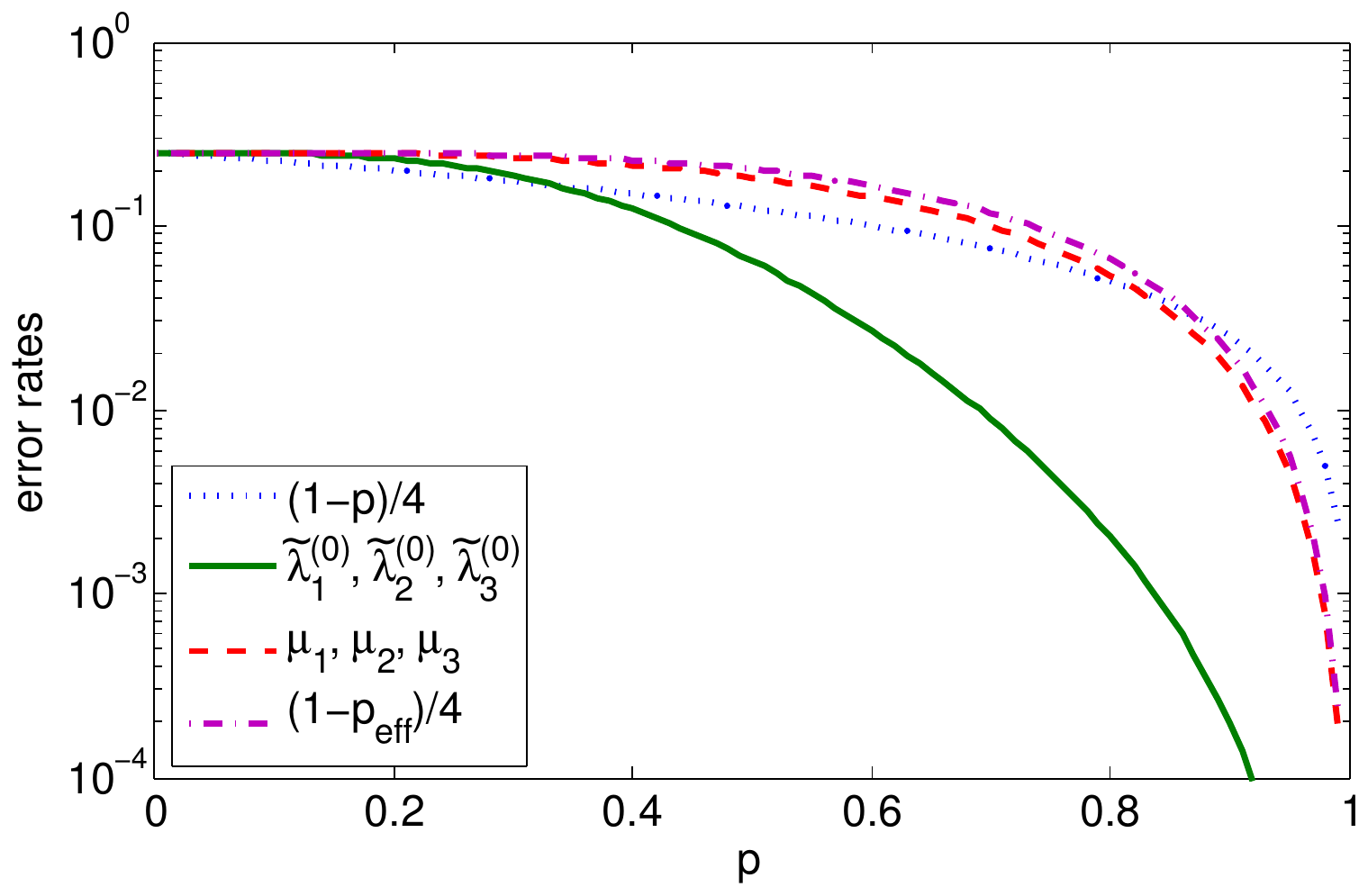}
 \caption{Effective error rates $\tilde{\lambda}_{i>0}^{(0)}$ and $\mu_{i>0}$ for the cluster-ring code with $m = 5$ and physical white noise. Since the effective noise is again of white noise structure, they equal for all $i$.  Both effective models give rise to a reduced effective error rate for small physical noise rates. The mean channel is compared to a simple estimate on the effective error rate $p_{\mathrm{eff}}$ (see text).}
 \label{fig:CR-x}
\end{figure}

The same is true for the mean noise channel (\ref{eq:4}), which is numerically calculated. However, the averaging diminishes the positive effect of the encoding. For both methods, see Fig.~\ref{fig:CR-x} for an illustration. Interestingly, the mean error rates $\mu_{i>0}$ are close to a simple estimate: If $p$ is the probability for no error on one such qubit, the overall probability to have at most one error equals $p_{\mathrm{eff}} = p^{5} + 5 p^4 (1-p)$. This can be seen on the logical level of having no error \cite{hein_entanglement_2005}). One finds that $\mu_{i>0} \approx (1-p_{\mathrm{eff}})/4$.

In summary, the repetition code and the cluster code thus lead to different effective noise descriptions. While we find an (enhanced) effective phase noise for the repetition code, the cluster-ring encoding leads to an effective white noise that is weaker than the initial noise.

\section{Lower bounds on entanglement measures}\label{sec:entanglement-measures}

The effective description of noise processes on the logical level can be applied to logical states $|\psi_{\mathrm{L}}  \rangle $. The effort in the analysis of their entanglement properties is then comparable with that of the unencoded state $| \psi \rangle $. As the effective noise description is based on local (in terms of the logical qubit system) operations, one always has lower bounds on entanglement lifetime and entanglement measures. 

In this section, we discuss the results of Sec.~\ref{sec:effective-noise-description} for the logical GHZ state, which is defined as
\begin{equation}\label{eq:ghz}
 \ket{\mathrm{GHZ}_{\mathrm{L}}}= \frac{1}{\sqrt{2}}\left(\ket{0_{\mathrm{L}}}^{\otimes N}+\ket{1_{\mathrm{L}}}^{\otimes N} \right).
\end{equation}
After some immediate implications based on already known results, we calculate lower bounds on the lifetime of distillable entanglement and the negativity for this state and the encodings of Sec.~\ref{sec:effect-noise-repet}.

\subsection{Implications for logical states $| \psi_{\mathrm{L}} \rangle $}
\label{sec:impl-logic-stat}

The GHZ state is an important instance of a genuine multipartite-entangled quantum state. It was shown (see, e.g., Ref.~\cite{aolita_scaling_2008}) that the GHZ state is very fragile with respect to phase noise. In fact, essentially all quantum properties decay exponentially fast in $N$. The choice of the repetition code seems therefore a bad choice to protect the \textit{logical} GHZ state against noise, because, on the logical level, the effective phase noise is even enhanced. On the other side, it was also found \cite{chaves_robust_2012} that the GHZ state in the eigenbasis of $\sigma_x$, [i.e., $  \left\{ \left| \pm \right\rangle = 1/\sqrt{2}(\left| 0 \right\rangle \pm \left| 1 \right\rangle ) \right\}$] is stable in the sense that the decay of, e.g., the negativity \cite{vidal_computable_2002} is asymptotically independent of $N$.

In the presence of Pauli noise that is undirected (e.g., white noise), a local rotation of the GHZ state does not lead to an improved stability. However, we can use a clever encoding $| 0_{\mathrm{L}} \rangle $ and $| 1_{\mathrm{L}} \rangle $ to protect the GHZ state from any Pauli noise. In analogy to Ref.~\cite{chaves_robust_2012}, one can define the eigenstates of the logical Pauli-$x$ operator $\sigma_x^L$ as the basis for the logical qubit, that is, we set
\begin{equation}
 \label{eq:3}
 \begin{split}
  \left| 0_{\mathrm{L}} \right\rangle  & \doteq \frac{1}{\sqrt{2}} \left( \left| 0 \right\rangle ^{\otimes m}+ \left| 1 \right\rangle ^{\otimes m} \right)\\
  \left| 1_{\mathrm{L}} \right\rangle  & \doteq \frac{1}{\sqrt{2}} \left( \left| 0 \right\rangle^{\otimes m} - \left| 1 \right\rangle^{\otimes m}  \right).\\
 \end{split}
\end{equation}
This was indeed done in Refs.~\cite{frowis_stable_2011,frowis_stability_2012}, where the increased stability of $| \mathrm{GHZ}_{\mathrm{L}} \rangle $ for the map (\ref{eq:3}) was shown. Due to its structure, this state was called concatenated GHZ state. Here, we refer to map (\ref{eq:3}) as the GHZ encoding.

Note that in Ref.~\cite{chaves_robust_2012} it was shown that all graph states can be made robust by a local transformation if the physical noise is a \textit{directed} Pauli noise. Together with our finding that depolarizing noise can effectively be converted into directed Pauli noise, this implies that any graph state can be stabilized with respect to any local Pauli noise by a repetition code with subsequent unitary rotation within the logical space $P_0$.

\subsection{Lifetime of Distillable Entanglement}

In the following, we are interested in the question up to which noise level a quantum state contains distillable entanglement. For simplicity, we restrict ourselves to the depolarizing channel (\ref{eq:13}) once more. Since noise channels are often associated with non-unitary time-evolution (e.g., $p = \exp(-\gamma t)$ with time $t \in \mathbbm{R}_{>0}$ and decoherence rate $\gamma \in \mathbbm{R}_{>0}$), the critical noise parameter $p_{\mathrm{crit}}$ corresponds here to the lifetime of entanglement. The smaller $p_{\mathrm{crit}}$, the larger is the lifetime of entanglement.

Clearly, one has to specify the entanglement measure. Here, we follow Ref.~\cite{hein_entanglement_2005} and calculate lower bounds on the ability to distill entanglement. There, a protocol has been derived that can be used for any graph state $| \mathrm{G} \rangle $. Here, we summarize this protocol and apply it to the logical GHZ state (\ref{eq:ghz}) using the effective noise channels (\ref{eq:eff-noise}). The basic idea is to distill maximally entangled states between any pair of qubits. Then, any other $N$-qubit state can be generated by local operations and classical communication. As long as the success probabilities for this protocol are nonzero, the encoded state contains entanglement. In this way, we find bounds on the lifetime of distillable entanglement.

As in Ref.~\cite{hein_entanglement_2005}, we assume that the logical state is some graph state consisting of $N$ qubits. Some noise process acts on the logical qubits, as effectively described by Eq.~(\ref{eq:eff-noise}) for $i=0$. If now all but two neighboring particles $(k,l) $ are measured with $\sigma_z^L$, the resulting state of the remaining two particles is --up to logical Z rotations-- described by
\begin{equation}
 \ket{\Phi}=\frac{1}{\sqrt{2}}\left(\ket{+_{\mathrm{L}}}\otimes\ket{0_{\mathrm{L}}}+\ket{-_{\mathrm{L}}}\otimes\ket{1_{\mathrm{L}}} \right) \in P_0^{\otimes 2}, 
\end{equation}
where $| \pm_{\mathrm{L}} \rangle = 1/\sqrt{2} (\left| 0_{\mathrm{L}} \right\rangle \pm \left| 1_{\mathrm{L}} \right\rangle )$.
Considering that the action of the Pauli noise process commutes (up to logical Z rotations) with $\sigma_z^L$ \cite{hein_entanglement_2005-1}, one easily sees that the evolution of the reduced two-body density operator $\rho_{kl}$ only depends on the action of the cp map $\tilde{\mathcal{E}}_j$ on the blocks $k$ and $l$ themselves and on their neighbors: 
\begin{equation}
 \rho_{kl}= \mathrm{Tr}_{1,\dots,N \backslash \left\{ k,l \right\}}\left[ \prod_{j\in I} \tilde{\mathcal{E}}_j  \left(\left| \mathrm{G} \right\rangle\!\left\langle
 \mathrm{G}\right| \right) \right]
\end{equation}
where $I= N_k \cup N_l \cup \{k\} \cup \{l\}$.

We invoke the PPT criterion \cite{peres_separability_1996,horodecki_separability_1996} to get a sufficient condition for distillability. Denoting the partial transposition of party $k$ by $T_k$, $\rho_{kl}$ exhibits distillable entanglement if
\begin{equation}\label{eq:PPT}
 \rho_{kl}^{T_k} \geq 0.
\end{equation}
In principle, this condition has to be shown for any pair of neighboring particles $k$ and $l$ of the logical graph states in order to prove distillable entanglement for the whole graph state. For some symmetric logical graph states, like the GHZ state (\ref{eq:ghz}), it is sufficient to study one arbitrary pair of neighboring particles, as all vertices of the graph are equivalent.

\begin{figure}[htb]
 \center
 \includegraphics[scale=0.65]{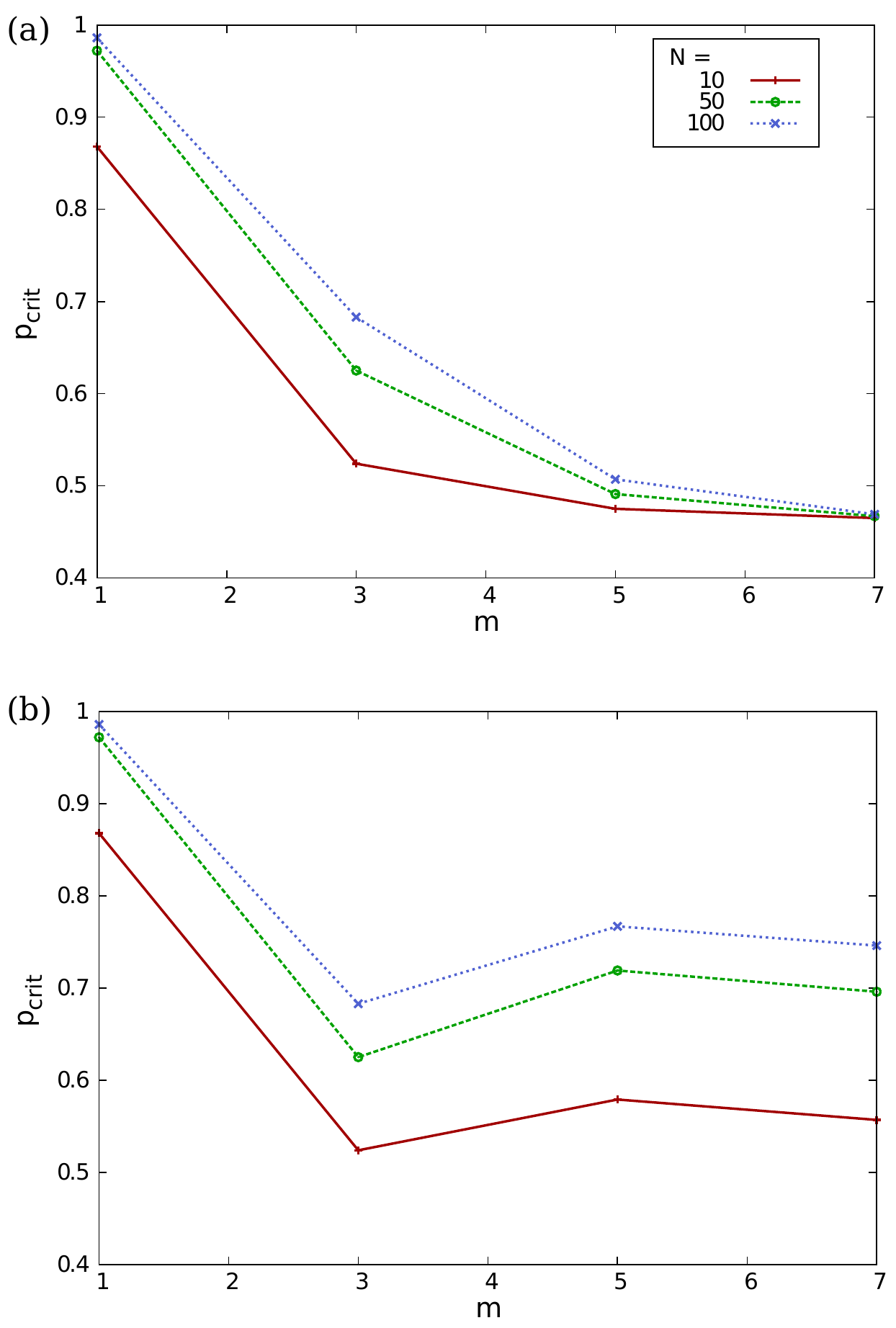}
 \caption{Lower bound on the lifetime of distillable entanglement of (\ref{eq:ghz}) (a) for GHZ encoding (\ref{eq:3}) and (b) for cluster-ring encoding (\ref{eq:cr-enc}) with different numbers of qubits $m$ used for the encoding and for different system sizes $N$. While for the GHZ encoding, the lower bound $p_{\mathrm{crit}}$ decreases by increases $m$, it does not decrease monotonically with $m$ for the cluster-ring encoding.}
 \label{fig:lifetime}
\end{figure}

We now use this distillation protocol to study the effect of encodings on the lifetime of distillable entanglement using effective noise channels. Since we investigate whether entanglement is present or not without quantifying it, it is sufficient to consider the projected effective channels $\tilde{\mathcal{E}}^{(0)}$ of Eq.~(\ref{eq:eff-noise}), which gives better bounds than the mean noise channel $\tilde{\mathcal{E}}_{\mathrm{mean}}$.
Our numerical results are presented in Fig.~\ref{fig:lifetime} (a) for the GHZ encoding (\ref{eq:3}) and in Fig.~\ref{fig:lifetime} (b) for the cluster-ring encoding (\ref{eq:cr-enc}). The critical value $p_{\mathrm{crit}}$ --below which there is no distillable entanglement present-- is compared to the number $m$ of physical qubits per logical block. We observe that the lower bound on the lifetime of distillable entanglement can be increased by choosing a larger number of qubits $m$ for the encoding. This is true for both types of encoding and for all tested system sizes $m$. Comparing both types of encoding reveals that the GHZ encoding is better suited to stabilize the logical GHZ state. That is, the lower bound on the lifetime of distillable entanglement is more increased if the GHZ encoding is chosen.

\subsection{Negativity}

As a further application of the effective noise channels, we investigate the negativity \cite{vidal_computable_2002} for encoded quantum states. The negativity is a measure for bipartite entanglement. Given any partition $a:b$ of the $N$-qubit state $\rho$, the negativity $\mathcal{N}$ is defined as
\begin{equation}
 \label{eq:6}
 \mathcal{N} \mathrel{\mathop:}= \frac{1}{2} \left( \lVert \rho^{T_a} \rVert _1 - 1\right).
\end{equation}
The function $\lVert \cdot \rVert_1$ is the trace norm and, for hermitian operators, nothing else than the sum of the moduli of their eigenvalues. A typical choice for a partition is $1:N-1$. In the case of logical encoding, we translate this to the partition of one logical block versus the remaining $N-1$ blocks.

As before, we consider the logical GHZ state (\ref{eq:ghz}) with the GHZ encoding and the cluster-ring encoding. In contrast to the analysis of the lifetime, we now have to consider all error subspaces of the encoding. Therefore, we make use of the mean noise channel (\ref{eq:4}).

\begin{figure}[htb]
 \center
 \includegraphics[scale=0.65]{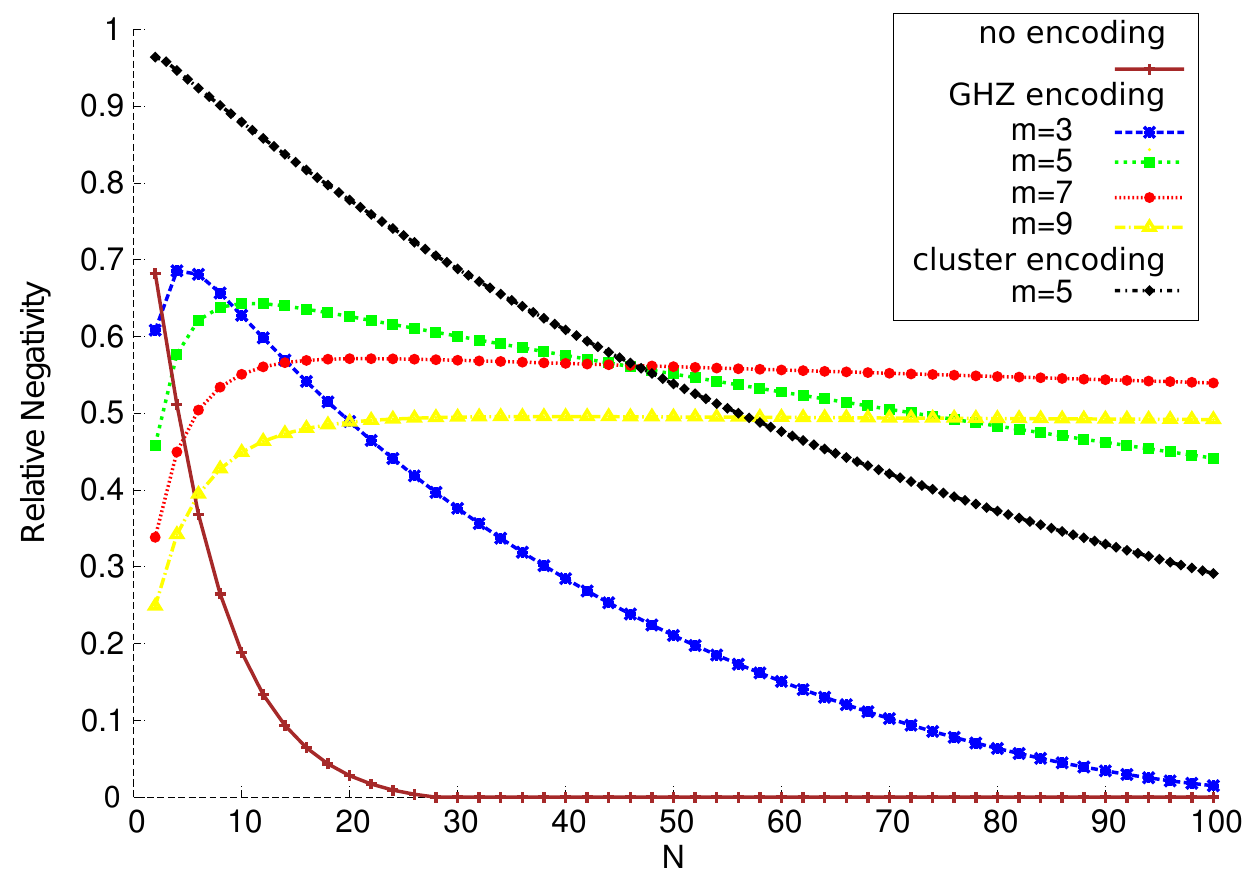}
 \caption{Lower bound on the negativity for the $N$-partite logical GHZ state with different $m$-qubit encoding and local white noise with $p = 0.95$. The considered splitting is $1:N-1$. For the GHZ encoding, even for large system sizes, $\mathcal{N}$ can be stabilized by choosing a large enough $m$. For the cluster-ring encoding, one observes better values for small $N$, while the decay rate with $N$ is larger compared to the GHZ encoding with $m= 5$.}
 \label{fig:Negativity}
\end{figure}

A general formula for $\mathcal{N}$ for the GHZ state under arbitrary Pauli noise is easy to derive (see, e.g., Eq.~5 in Ref.~\cite{chaves_robust_2012}). 
It is then straightforward to calculate the negativity even for large system sizes $N$. With the formalism of the mean noise channel (\ref{eq:4}), we easily find lower bounds on the negativity for the GHZ encoding and the cluster-ring encoding. These findings are presented in Fig.~\ref{fig:Negativity}.

As mentioned before, the logical GHZ state $| \mathrm{GHZ}_{\mathrm{L}} \rangle $ with GHZ encoding (\ref{eq:3}) was was already studied in Refs.~\cite{frowis_stable_2011,frowis_stability_2012}. There, a full and computationally more costly analysis of the negativity was done. Comparing Fig.~\ref{fig:Negativity} with Fig.~4 (b) of Ref.~\cite{frowis_stability_2012}, we see that qualitatively equivalent results are found. In particular, the increment of $m$ leads to an exponential stabilization of the rate with which $\mathcal{N}$ decays with $N$.

By comparing the Negativity of $| \mathrm{GHZ}_{\mathrm{L}} \rangle $ for both encodings with $m=5$, we see that for large system sizes $N>N_{\mathrm{crit}}\approx 46$ the GHZ encoding is better suited to stabilize the logical qubit state. The threshold value $N_{\mathrm{crit}}$ dependents on the value of $p$. What is not shown here is that $N_{\mathrm{crit}}$ decreases if $p$ becomes smaller.

\section{Concatenation of Encodings}
\label{sec:conc-encod}

A question that arises immediately in the discussion of encodings is whether it is favorable to use several levels of encodings. A simple, effective and well known instance of an error correcting code is the nine-qubit Shor code \cite{shor_scheme_1995}. It is defined as
\begin{equation}
 \label{eq:7}
 \begin{split}
  \left| 0_{\mathrm{L}} \right\rangle &\doteq  \left[ \frac{1}{\sqrt{2}}\left(
      \left| 0 \right\rangle ^{\otimes 3} + \left| 1 \right\rangle
      ^{\otimes 3} \right) \right]^{\otimes 3},\\
    \left| 1_{\mathrm{L}} \right\rangle &\doteq  \left[ \frac{1}{\sqrt{2}}\left(
      \left| 0 \right\rangle ^{\otimes 3} - \left| 1 \right\rangle
      ^{\otimes 3} \right) \right]^{\otimes 3}.
 \end{split}
\end{equation}
This code can be seen as a GHZ encoding [cp.~to Eqs.~(\ref{eq:9}) and (\ref{eq:3})] with $m = 3 $ that is repeated three times. On the lower level --the GHZ encoding-- one bit-flip error can be corrected, while on the higher level --the repetition of the GHZ encoding-- one can correct a single phase-flip error. With the effective channels derived in this work, it is easy to understand why the Shor code works; in particular, why the increased sensitivity on the lower level to phase noise does not spoil the ability to correct it on the higher level (and similar for flip errors). The reason is that the suppression of phase errors of the higher scale is larger than the increment of the sensitivity to this kind of error on the lower level.

The idea of Shor's code can be generalized to more levels and to more qubits per level. Consider, for example, the following generalization of the Shor code to 
\begin{equation}
 \label{eq:7a}
 \begin{split}
  \left| 0_{\mathrm{L}} \right\rangle &\doteq  \left[ \frac{1}{\sqrt{2}}\left(
      \left| 0 \right\rangle ^{\otimes m_1} + \left| 1 \right\rangle
      ^{\otimes m_1} \right) \right]^{\otimes m_2},\\
    \left| 1_{\mathrm{L}} \right\rangle &\doteq  \left[ \frac{1}{\sqrt{2}}\left(
      \left| 0 \right\rangle ^{\otimes m_1} - \left| 1 \right\rangle
      ^{\otimes m_1} \right) \right]^{\otimes m_2},
 \end{split}
\end{equation}
where $m_1,m_2 \in \mathbbm{N}$. This code can correct up to $\lfloor m_2/2 \rfloor$ phase errors and up to $\lfloor m_1/2\rfloor$ flip errors and has, for instance, been used to investigate the non-zero capacity of depolarizing noise, see Refs.~\cite{divincenzo_quantum-channel_1998,smith_degenerate_2007,fern_lower_2008,chen_graph-state_2011}. Here, we study two question concerning this noise model. First (which is related to the non-zero capacity problem), up to which noise level (given by the parameter $p$) are the logical error rates smaller than the physical ones? The second question is with respect to entanglement: Does this concatenated structure of the logical states give rise to a better distillability or a longer lifetime of entanglement of the logical GHZ state (\ref{eq:ghz})?

The effective noise description that we introduced in this work offers a handy tool to study the effects of multiple concatenation levels, as the effective description can be calculated successively for each of the levels. One starts with the lowest level (the physical qubits) to derive its effective noise description. This effective noise can then be applied on the second-last level to derive an effective description for this level. This protocol is repeated for any level of concatenation. The effective noise description for the highest level of concatenation can then be applied to the logical qubit.

\begin{figure}[htbp]
 \centerline{\includegraphics[width=\columnwidth]{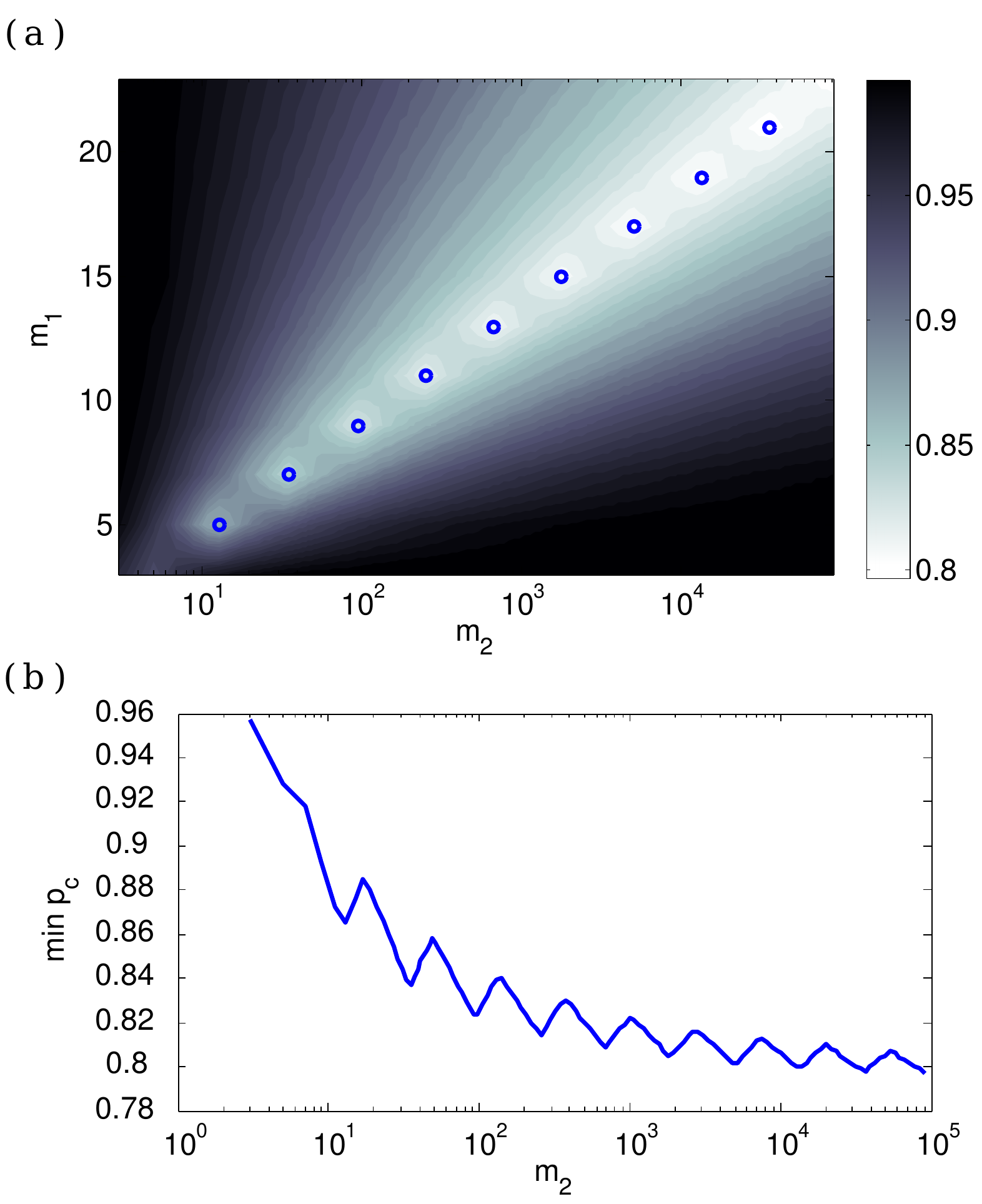}}
 \caption[]{\label{fig:shor} (a) Critical noise rate $p_{\mathrm{c}}$ above with the physical error rates are higher than the logical ones using the generalized Shor code (\ref{eq:7a}). For any pair $(m_1,m_2)$, $p_{\mathrm{c}}$ is color coded: Lighter color means lower (i.e., better) critical rates. The blue circles mark local minima [see also (b)]. For very large $(m_1,m_2)$, a critical noise parameter below 0.8 can be reached. Note that there is an exponential relation between $m_1$ and $m_2$. (b) Minimization of $p_{\mathrm{c}}$ with respect to $m_1$. The local minima are correspond to the circles in (a).}
\end{figure}

Coming back to the first question concerning error rates, we numerically compare the mean noise channel to the physical noise rates for a given pair $(m_1,m_2)$ and a given $p$. For very small $p$ (i.e., large error rates), the code always gives worse rates than the physical channel. Hence there exists a critical rate $p_{\mathrm{c}}$, where the logical error rates are smaller than the physical ones for any $p\geq p_{\mathrm{c}}$. In Fig.~\ref{fig:shor}, the result of this search is presented. We find that the smallest $p_{\mathrm{c}}$ are achieved by using an exponential relation  $m_2 \propto \exp(c m_1)$, with $c$ a positive scalar.

Concerning the second question, a study of the logical GHZ state with encoding (\ref{eq:7a}) shows that, for large system sizes and strong noise processes, one single concatenation level increases the lifetime of distillable entanglement of the  logical $N$-particle GHZ state more than two level of concatenation.

Finally, we remark that the presented method can be also applied to combinations of different codes. This could lead to new useful codes as one can quickly check their merit. 

\section{Summary}\label{sec:outlook}

Under the influence of noise, it is important to study and understand the stability or fragility of entanglement in multipartite quantum states. A general strategy to protect entanglement and other crucial features of a quantum system is to introduce a redundant encoding of quantum information. However, quantitative answers to the question how large the merit of a specific encoding is nontrivial. This is mainly due to the fact that the encoding leads to an increased complexity of the problem.

In this paper, we presented a simple scheme to reduce this complexity to the original problem of the unencoded system. The idea was to replace the noise map acting on the physical qubits that constitute the logical qubit by a map acting only on the logical (two-dimensional) subspace. We presented two possible realizations of such a reduction. For qualitative questions on entanglement (like the lifetime of distillable entanglement), it is sufficient to project the noisy logical qubit to its original subspace and consider the effective map after renormalization. To estimate the amount of entanglement, it is necessary to consider deterministic protocols. We therefore consider the effective error rates from all projections onto the different subspaces defined through the error code. The sum of these single contributions weighted by their probability of appearance gives rise to a mean noise channel. Applying this channel to an encoded system is then an upper bound of the effect of the physical noise but the complexity of the problem is (drastically) reduced. In addition, the effective noise channels can give rise to a simple and intuitive explanation why and how error codes work.

We examined these ideas to two different error codes: the repetition and the cluster-ring code. The repetition code transforms (undirected) white noise to directed effective Pauli noise that is stronger than the original noise. However, certain states like the GHZ state can better cope with this noise if the direction is properly chosen (which is done through the choice of the basis within the logical subspace). On the other side, the cluster-ring encoding transforms white noise to white noise on the logical level with reduced error rates if the physical noise is weak.

For the logical GHZ state, we have shown that the lifetime of distillable entanglement under white noise can be increased by the right choice of the basis of the repetition code (we called it the GHZ encoding), while the cluster-ring code gives only moderate improvement for the lifetime. We see that in this case it is not so important to be able to correct all occurring errors since some of them do not influence the distillability as much as others. The same is true for entanglement measures like the negativity. Here we found that the GHZ encoding again leads to larger entanglement than the cluster-ring encoding for large system sizes.

Finally, we demonstrated how easy our formalism can be used to analyze concatenated codes like the Shor code or its generalizations. On every level of the concatenation, one can apply the idea of the effective noise channel in order to estimate the merit of the concatenation.

In summary, we proposed a powerful tool to investigate entanglement properties of encoded systems. We are confident that the simplicity of our approach leads to new insights in the mechanisms of error codes and to new ideas how to design novel codes.

\begin{acknowledgments}
 This work was supported by the Austrian Science Fund (FWF): Grants No. P24273-N16 and No. SFB F40-FoQus F4012-N16.
\end{acknowledgments}

\bibliographystyle{apsrevMOD}
\bibliography{EffectiveNoise}
 
\end{document}